\documentclass[referee]{raa}            % referee version: for submission

\usepackage{graphicx,times}             %for PS/EPS graphics inclusion, new
\usepackage{booktabs}
\begin{document}

   \title{New Vacuum Solar Telescope and Observations with High Resolution
%\footnotetext{$*$ Supported by the National Natural Science Foundation of China.}
}
%   \subtitle{I. Place Your Subtitle Here}

   \volnopage{Vol.0 (200x) No.0, 000--000}      %%preserved for Editor. DOn't remove!
   \setcounter{page}{1}          %%starting page, preserved for Editor. DOn't remove!

   \author{Zhong Liu
      \inst{1}
   \and Jun Xu
      \inst{1}
   \and Bo-Zhong Gu
      \inst{2}
   \and Sen Wang
      \inst{3}
   \and Jian-Qi You
      \inst{4}
   \and Long-Xiang Shen
      \inst{3}
   \and Ru-Wei Lu
      \inst{1}
   \and Zhen-Yu Jin
      \inst{1}
   \and Lin-Fei Chen
      \inst{1}
   \and Ke Lou
      \inst{1}
   \and Zhi Li
      \inst{1}
   \and Guang-Qian Lu
      \inst{1}
   \and Zhi Xu
      \inst{1}
   \and Chang-Hui Rao
      \inst{5}
   \and Qi-Qian Hu
      \inst{2}
   \and Ru-Feng Li
      \inst{1}
   \and Hao-Wen Fu
      \inst{1}
   \and Men-Xian Bao
      \inst{1}
   \and Ming-Chan Wu
      \inst{1}
   \and Bo-Rong Zhang
      \inst{1}
   }

   \institute{Yunnan Observatories, Chinese Academy of Sciences, Kunming 650011, China; {\it lz@ynao.ac.cn}\\
        \and
             Nanjing Institute of Astronomical Optics \& Technology, National Astronomical Observatories, Chinese Academy of Sciences, Nanjing 210042, China\\
        \and
             National Astronomical Observatories, Chinese Academy of Sciences, Beijing 100012, China\\
        \and
             Purple Mountain Observatory, Chinese Academy of Sciences, Nanjing 210008, China\\
        \and
             Institute of Optics and Electronics, Chinese Academy of Sciences, Chengdu 610209, China\\
   }

   \date{Received~~2013 month day; accepted~~2013~~month day}

\abstract{ The New Vacuum Solar Telescope (NVST) is a 1 meter vacuum solar telescope that aims to observe the fine structures on the Sun. The main tasks of NVST are high resolution imaging and spectral observations, including the measurements of solar magnetic field. NVST is the primary ground-based facility of Chinese solar community in this solar cycle. It is located by the Fuxian Lake of southwest China, where the seeing is good enough to perform high resolution observations. In this paper, we first introduce the general conditions of Fuxian Solar Observatory and the primary science cases of NVST. Then, the basic structures of this telescope and instruments are described in detail. Finally, some typical high resolution data of solar photosphere and chromosphere are also shown.
\keywords{telescopes --- instrumentation: adaptive optics --- instrumentation: polarimeters --- instrumentation: spectrographs --- techniques: high angular resolution --- Sun: photosphere --- Sun: chromosphere --- Sun: magnetic fields }
}

   \authorrunning{Z. Liu, J. Xu \& et al. }            %author_head in even pages
   \titlerunning{New Vacuum Solar Telescope and Observations with High Resolution }  % title_head in odd pages

   \maketitle
%% The author head (on even pages) and the title head (on odd pages) will be
%% automatically extracted from \author{} and \title{}. Whenever the title is too long,
%% you will be asked to supply a shorter one by inserting either \authorrunning{} or
%% \titlerunning{} before \maketitle. Anyway, you can specify your own heads.
%%
%%
%% Note: In the following text body of your manuscript, please note several differences from
%%       other major journals:
%% (1) \subsection{Please Capitalize the First Letter of Each Notional Word in Subsection Title}
%% (2) Please Capitalize the First Letter of Each Notional Word in all tables' captions

%
%________________________________________________ sections below
%
\section{Introduction}           %% first-level sections will be auto-capitalized
\label{sect:intro}

\begin{figure}
   \centering
   \includegraphics[width=\textwidth, angle=0]{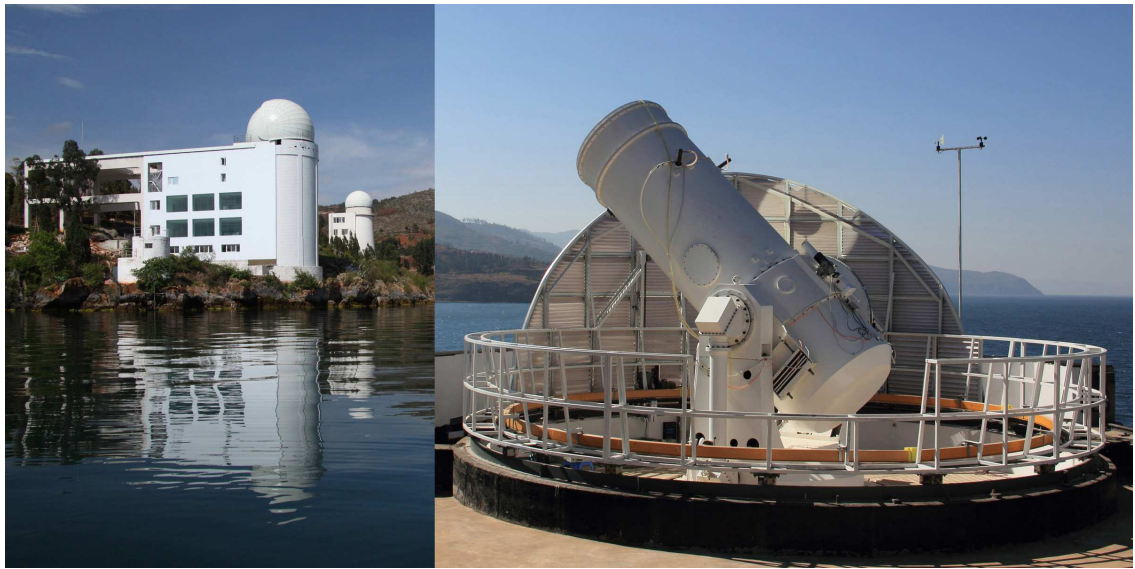}
   \caption{Building (left), telescope and the wind screen (right) of NVST.}
   \label{Fig1}
   \end{figure}

\begin{figure}
   \centering
   \includegraphics[width=\textwidth, angle=0]{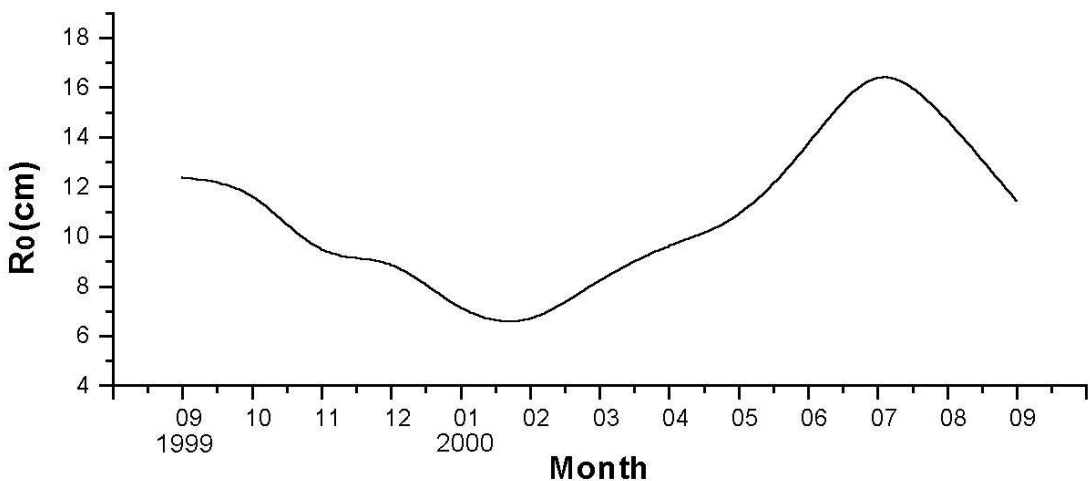}
   \caption{Variations curve of the seeing (r$_{0}$) at FSO in the period from September 1999 to September 2000.}
   \label{Fig2}
   \end{figure}

The New Vacuum Solar Telescope (NVST) is a vacuum solar telescope with 985 mm clear aperture. It is the primary observation facility of the Fuxian Solar Observatory (FSO). The location of FSO is \ensuremath{24^{\circ}}34\ensuremath{'}48\ensuremath{''}N and \ensuremath{102^{\circ}}57\ensuremath{'}01\ensuremath{''}E, the northeast side of the Fuxian Lake (Figure \ref{Fig1}), with the altitude of 1,720 m above the sea level. The average seeing (Fried parameter, r$_{0}$) of FSO obtained in the period from 1998 to 2000 was about 10 cm (Figure \ref{Fig2}). The sunshine duration of FSO is about 2,200 hours per year. The average wind velocity is 6 m s$^{-1}$, more than 75\% wind around FSO is from the lake and toward the telescope (\cite{lou01}). The weather parameters were measured by an automatic meteorological station. The seeing parameters include the scintillation and the Fried parameter, and were measured by the scintillometer and the Solar Differential Image Motion Monitor (SDIMM) that was first developed by the FSO team (\cite{liu01}; \cite{becker03}).

%      A figure as large as the width of the column
%-------------------------------------------------------------
%
%      A figure as large as the width of the column
%-------------------------------------------------------------

The early science cases of NVST were focused on the spectrum observation because NVST was originally proposed mainly as the ground-based large scale spectrometer of the Chinese one meter Space Solar Telescope (\cite{deng09}). The original primary working mode of NVST is multi-bands spectrum observation, including the Stokes parameter measurement of the solar magnetic sensitive lines. The good seeing of FSO encourages the FSO team to expand the scientific goals to high resolution observations in order to cover more topics related to open and hot issues. Now, the scientific goals of the telescope include observing the Sun with very high spatial and spectral resolutions in the wavelength range from 0.3 to 2.5 micron; detecting small scale structures and the fine details in the evolution of the solar magnetic field and their coupling to the plasma; investigating the energy transfer, storage and release in the solar atmosphere, such as the corona heating, the triggering of the solar eruption, and the other key questions regarding solar activities. Now, as the primary optical and near infrared solar telescope of Chinese solar community (\cite{fang11}; \cite{wang13}), NVST is required to undertake more science and technology tasks, including the key experiments of the next generation solar telescopes (\cite{liu12}).

In the next section, we are briefly describing the basic structure of NVST. The instrumentation of the telescope will be introduced in Section 3. We display some high resolution observational data obtained by NVST in Section 4.

%% Authors can give a citation as 'Michel et al. 1992'.
%% You may also use \cite, \citep and \citet for citation, and use Table~1 or Figure~1
%% and so forth. Using \ref and \label for cross-references of Tables/Figures
%% is a good way in adjusting/adding/removing text, tables or figures.

\section{Basic Structure of NVST}
\label{sect:Basic}
\subsection{Mechanism and building}
\begin{figure}
   \centering
   \includegraphics[width=\textwidth, angle=0]{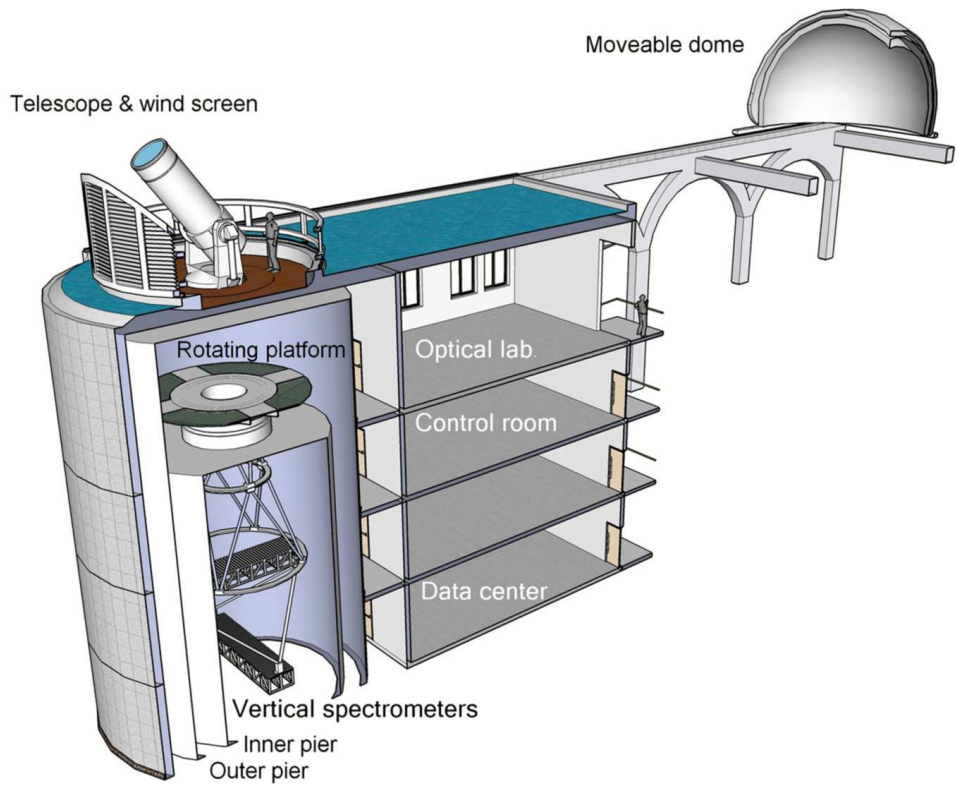}
   \caption{3-D sketch of the whole NVST.}
   \label{Fig3}
   \end{figure}

The whole building of NVST is a complex system and mainly includes a vacuum telescope, an instrument platform, vertical spectrometers and the other necessary assistant equipments, such as the wind screen and the moveable dome. Figure \ref{Fig3} is the 3-D sketch of the whole building. The telescope is installed on the front top of a 16 meter high building. The building is constructed on a big rock, very close to the lake (Figure \ref{Fig1} left). The roof of the building is designed to be a shallow pool filled with water to cool the floor. In most cases, the telescope works on the open air in order to keep the good local seeing. The moveable dome will open and move to another side of building while telescope is working. A louvered wind screen fences telescope against the wind and guide the wind toward the floor, reducing the near ground turbulence. The wind screen can move and rise automatically according to the wind direction and the different attitudes of telescope. The telescope and the instrument platform are individually installed on two independent piers to avoid the vibration crosstalk. The dome and the wind screen are located on the building directly.

The mounting system of NVST is an alt-azimuth structure (Figure \ref{Fig1}, right). The telescope should rotate on the altitude axis and the azimuth axis against the earth's rotation and tracking the Sun. The alt-azimuth mounting means stabilization and small wind resistance as well as the observation blind zone and the non uniform image rotation. The blind zone near zenith position is caused by the accelerating limit of the azimuth rotation of the telescope. The blind zone of NVST is less than 2 degree from zenith point, considering the latitude of the Tropic of Cancer is 23.5$^\circ$ and the latitude of FSO is 24.5$^\circ$N, there is no blind area for solar observation except in several days around the summer solstice. The image rotation is caused by the asymmetric turning of optical axis and the relatively rotating between mirrors. In order to eliminate the image rotation, the instrument platform and the vertical spectrometers should rotate round the primary optical axis with non uniform velocity. The rotations of telescope and instrument platform are driven by several fine electric motors via the friction drive devices.

%      A figure as large as the width of the column
%-------------------------------------------------------------

The pointing accuracy of NVST is high enough to point any region on solar disk in several arc-seconds. The tracking accuracy of NVST is 0.3 arc-second (RMS). The pointing and tracking accuracy are maintained by two control loops. The sensors of the outer control loop are two high accuracy angle encoders installed on the altitude axis and the azimuth axis. The inner active control loop is an optical Auto Guide System (AGS). The main part of this AGS is an individual small guide telescope banding on the telescope tube. The sensor of the AGS is a 4k by 4k CMOS camera. NVST can steadily track the sun in several hours once AGS is operated. The key functional parameters of NVST are displayed in table 1.

%               one-column-spanning table
%________________________________________ Table 2: Use_of_the routines
\begin{table}
\begin{center}
\caption[]{ Key parameters of NVST}
\label{Tab:tab1}

%%Please Capitalize the First Letter of Each Notional Word in table's caption

 \begin{tabular}{ll}
  \hline\noalign{\smallskip}
     Clear aperture (mm)                        & 985 \\
     Field of view (arc-minute)                 & 3 \\
     Focal length (EEFL) at F3 (m)              & 45\\
     Spectral range (micron)                    & 0.3 $\sim$ 2.5  \\
     Tracking accuracy (arc-second)	            & $<$ 0.3  \\
     Pointing accuracy (arc-second)	            & $<$ 5    \\
     Image quality (80\% EE of PSF, arc-second)	& $<$ 0.4 at 550 nm  \\
     Blind area (degree)                        & $<$ 2 \\
     Vacuum degree inside telescope tube (Pa)   &	$<$ 100  \\
  \noalign{\smallskip}\hline
\end{tabular}
\end{center}
\end{table}

\subsection{Optical system of NVST}

\begin{figure}
   \centering
   \includegraphics[width=\textwidth, angle=0]{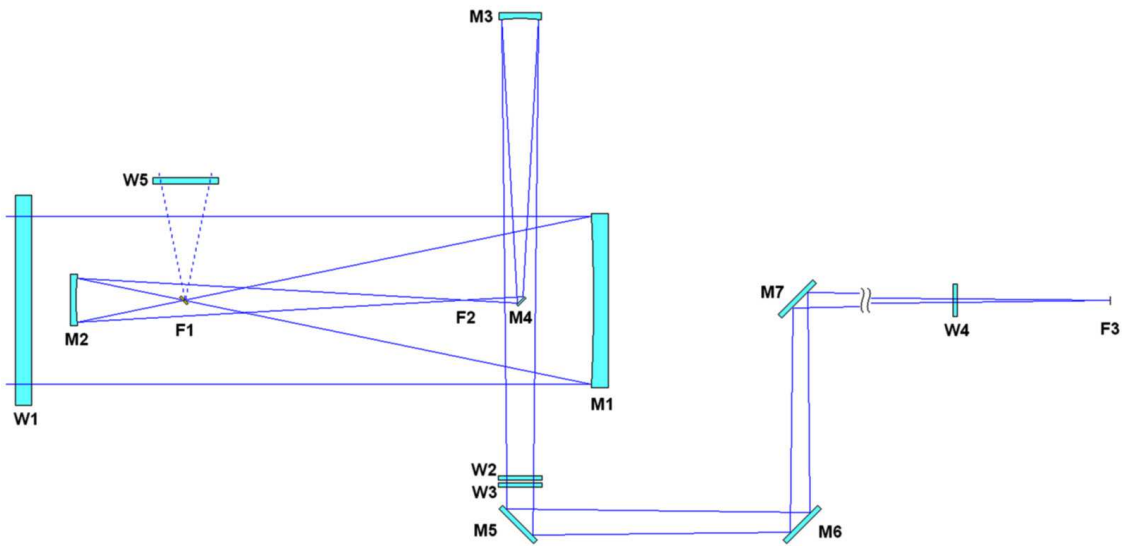}
   \caption{3-D sketch of the whole NVST.}
   \label{Fig4}
   \end{figure}

NVST is designed to be a vacuum telescope in order to reduce the turbulence in the system. Figure \ref{Fig1} right and Figure \ref{Fig4} display the telescope and its optical layout. An optical window (W1) with 1.2 meter diameter is placed on the top of vacuum tube to keep the air pressure inside the tube lower than 70 Pa. The optical system after W1 is a modified Gregorian system with an effective focal length of 45 m. The primary mirror is a parabolic imaging mirror with the clear aperture of 985 mm. There is a 3 arc-minute field diaphragm (heat stop) at the primary focus (F1) to prevent the extra energy entering the system afterward. Useless light is reflected from the system through another vacuum window (W5) on the side of the vacuum tube. The secondary mirror (M2) converges light to F/9 beam and focuses the light beam at the secondary focus (F2) where the calibration unit of polarization is installed. The M4 is a small flat mirror to reflect light toward horizontal direction. As the third imaging mirror, M3 converges light to the third focus (F3) after three flat reflectors (M5-M7). The mirror M3 is also the focusing mirror of the whole system. Table 2 displays more information of each optical component of NVST.

%      A figure as large as the width of the column
%-------------------------------------------------------------

The vacuum system consists of two vacuum tubes as the telescope should rotate on altitude axis and azimuth axis. These two vacuum tubes are separated by two vacuum windows (W2 \& W3). Most of the big optical components, including W1 through M4, are installed in the primary vacuum tube while the Mirrors M5, M6 and M7 are installed in the second vacuum tube. After going through all the mirrors and windows, the light from the Sun is focused and guided to a rotating instrument platform with the diameter of 6 meter, on which all imaging instruments, including adaptive optics system, were installed on this platform. The photosphere and the chromosphere can be observed with high spatial resolution by these imaging instruments. Two vertical grating spectrometers with different dispersion power are placed in a vertical hanging bracket (Figure \ref{Fig3}) which is hanged below the instrument platform and rotates along with the platform to eliminate the image rotation.

%               one-column-spanning table
%________________________________________ Table 2: Use_of_the routines
\begin{table}
\begin{center}
\caption[]{ Key parameters of optical components}
\label{Tab:tab2}

%%Please Capitalize the First Letter of Each Notional Word in table's caption

 \begin{tabular}{llll}
  \hline\noalign{\smallskip}
  Component	Size & Curvature(mm) and Shape & Material & Description\\
  \hline\noalign{\smallskip}
     W1	        & $\Phi$1,200, infinity, Flat    & BaK7          & Vacuum window \\
     M1         & $\Phi$985, R4,800, Paraboloid  & Glass-ceramic & Primary Mirror \\
     M2	        & $\Phi$258, R980, Ellipsoid	& Glass-ceramic	& Secondary Mirror \\
     M3      	& $\Phi$225, R3,238, Ellipsoid   & Glass-ceramic & Focusing Mirror  \\
     M4$\sim$M7	& $\ast$,infinity,Flat	        & Glass-ceramic	& Reflective mirrors  \\
     W2$\sim$W5	& $\ast$,infinity,Flat          & BaK7	        & Vacuum windows    \\
  \hline\noalign{\smallskip}
  \multicolumn{4}{l}{$\ast$ The sizes of these optical components are slightly bigger than the requirements.}\\
  \noalign{\smallskip}\hline
\end{tabular}
\end{center}
\end{table}

The thermal control system of NVST consists of two parts, a water cooling system and a set of heat pipes installed between F1 diaphragm and the wall of the vacuum tube. The residual heat is still harmful to F1 diaphragm and the infrared observations although most of the extra energy could be reflected out from the system. The heat pipes bring the residual heat to an exchanger connected to the water cooling system. A circular water cooling tunnel around W1 is designed to reduce the image degradation caused by the radial temperature gradient of W1. Another method to decrease the influence of the radial temperature gradient is to enlarge the diameter of W1. Therefore, the diameter of W1 is designed to be 1,200 mm, more than 200 mm larger than the diameter of the primary mirror. An air knife can be also installed at the edge of W1 as an optional device to keep the good mirror seeing.

\section{Instrumentations of NVST}
\label{sect:instr}

\begin{figure}
   \centering
   \includegraphics[width=\textwidth, angle=0]{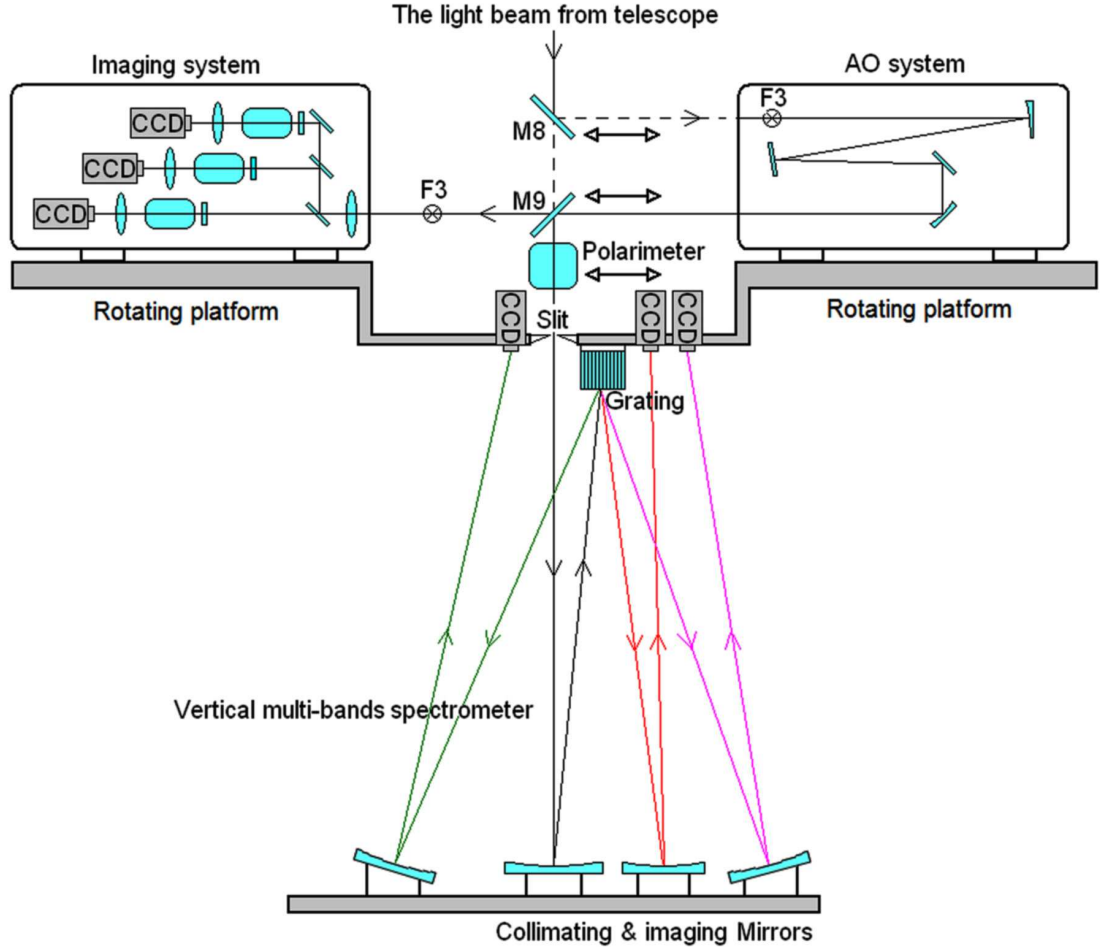}
   \caption{Schematic descriptions of instruments arrangement of NVST (Not in scale).}
   \label{Fig5}
   \end{figure}

The instruments equipped for NVST are designed for realizing observations and studies of the scientific goals as expected. They are located either on the rotating instrument platform or on the frames of the vertical hanging bracket below the telescope (Figure \ref{Fig3}). These instruments basically consist of three groups according to their functions and working modes, and the arrangement for them in space is sketched in Figure \ref{Fig5}. The first group includes adaptive optical (AO) system and polarization analyzer (PA). The AO system is the front equipment installed on the rotating platform before other
instruments, and PA is placed right before the slit of vertical spectrometers. The second group consists of all imaging instruments placed on the rotating platform. The third group includes two vertical grating spectrometers placed in the hanging bracket below the rotating platform.

%      A figure as large as the width of the column
%-------------------------------------------------------------

Each instruments could work either individually or as a subsystem of the whole system, which is connected to the telescope via mirrors M8 and M9 (Figure \ref{Fig5}). Here, M8 is a reflect mirror, and M9 is a beam splitter with high reflectance and low transmission. Different observing modes are realized by various combinations of M8 and M9. Light beam from the telescope is first reflected into the AO system by M8. Then, it is sent to different instruments depending on the purposes of observers after going through the AO system. The light beam does not go through M9 if the purpose of observations is just imaging only, and it will enter the imaging system directly after leaving the AO system. For spectrum observations, the beam splitter, M9, will be inserted into the optical path after the AO system. In this case, most photons are reflected down to the spectrometers, and the other photons will enter the imaging system for slit position monitoring. For observations without AO, M8 is removed from the optical path. Replacing M9 with a reflector, all the photons are sent to the imaging system. If M8 and M9 are both removed from the optical path, the photons may enter the spectrometers directly.

%      A figure as large as the width of the column
%-------------------------------------------------------------
   \begin{figure}
   \centering
   \includegraphics[width=\textwidth, angle=0]{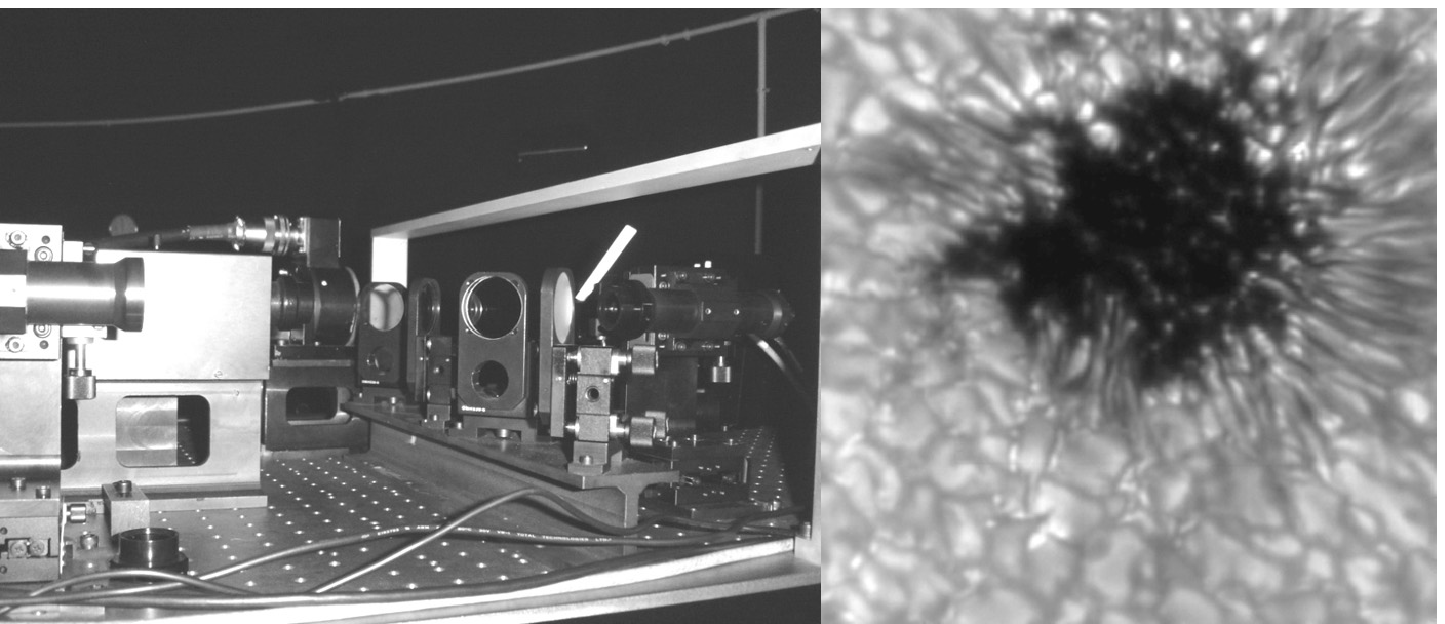}
   \caption{The low order AO system and the observation result obtained at 7,058 {\AA}.}
   \label{Fig6}
   \end{figure}

\subsection{Adaptive optical system of NVST}

The present AO system of NVST is a low order system with only 37 actuators(\cite{rao12}). It is placed before the other instruments but after F3 (Figure \ref{Fig5}). A Shack-Hartmann wave-front sensor with 37 sub-apertures is used to detect the wave-front by observing granulations or the other obvious small photosphere structures, such as small sunspots or pores. The frame rate of wave-front sensor is up to 800 fps (frames per seconds). The deformable mirror is a 40 mm glass mirror actuated by 37 piezoelectric transducer (PZT) actuators. The first 5 orders of the Zernike aberrations that include the front 20 modes could be efficiently corrected by this system. The Strehl ratio of corrected image is better than 0.5 when seeing is better than or equal to 10 cm. The left panel in Figure \ref{Fig6} displays some elements of the system, and that at right shows the observational results obtained at 7,058 {\AA} in the case of good seeing.

\subsection{Polarization measurement}

The basic approach to magnetic field detecting of NVST is the polarization measurement according to the Zeeman effect. The main structure of the telescope before M4 is strictly symmetrical, both optically and mechanically, in order to reduce extra instrument polarization or polarization crosstalk. For example, supporting spiders of the second mirror and the heat stop are cross type rather than the simpler and easy adjusting tripod style. The original design of PA for NVST was a fast modulation system with liquid crystal wave plates (\cite{xu06}). It was proposed to install close to F2 for high precision measurement of the solar polarization. Eventually, it turned out that this PA did not fit the space in the optical system of NVST assigned to it in the original design. It has now been replaced by a rotating modulation system with the classical wave plate. The present PA is placed before the slit of the spectrometer while the calibration unit is still installed close to F2. The whole polarization system is achromatic around both 5,000 and 10,000 {\AA}. Combining with the spectrometers, the system could conduct measurements of the Stokes parameters with high accuracy in both optical and near infrared bands, for example at 5,324 and 10,830 {\AA}. The PA and the calibration unit can be moved in or moved out from the optical path simultaneously or separately depending on the observing modes. The anticipated accuracy of polarization measurement is expected to be better than 5$\times$10$^{-3}$ after the ongoing installation and the careful calibration are completed.

\subsection{Imaging system}
Observing the fine structures in both the photosphere and the chromosphere is the main scientific goal of NVST. The imaging system is placed on the rotating platform with 6 meter diameter. Its main structure is a multi-channel high resolution imaging system and consists of one chromospheres channel and two photosphere channels. The band for observing the chromosphere is H$\alpha$ (6,563 {\AA}). The bands for observing the photosphere are TiO (7,058 {\AA}) and G-band (4,300 {\AA}), respectively. The H$\alpha$ filter is a tunable Lyot filter with the bandwidth of 0.25 {\AA}. It can scan spectrum in $\pm$5 {\AA} range with 0.1 {\AA} step length. Connected to the optical splitters, all the channels could observe and record the images synchronously. This means that the fine structures and their evolution in the photosphere and the chromosphere could be observed at the same time.

\begin{figure}
   \centering
   \includegraphics[width=\textwidth, angle=0]{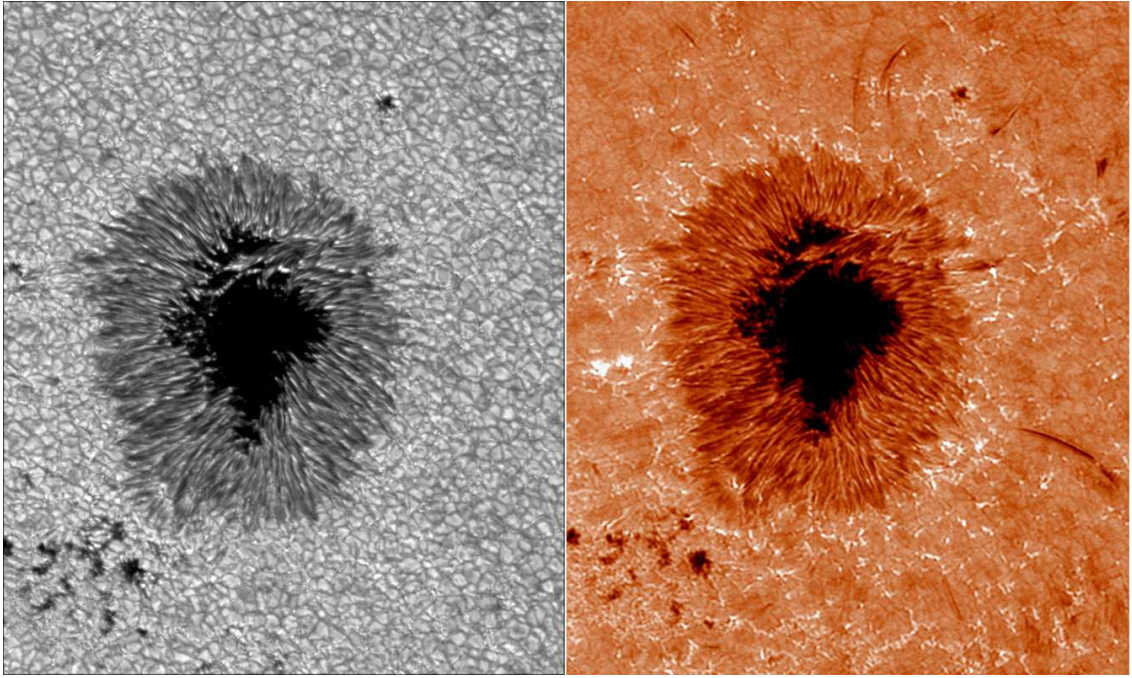}
   \caption{High resolution reconstructed images of the photosphere (7,058 {\AA}, left) and the chromosphere (H$\alpha$-0.8 {\AA}, right) of the AR11598.}
   \label{Fig7}
   \end{figure}

The whole system can work independently or work as an AO multi-channel imaging system by combining with the AO system. In a good seeing condition, without AO, the resolution of reduced photosphere images at G-band could almost reach 0.1 arc-second after reconstructing with statistical high resolution algorithm. The time cadence of the reconstructed images is about 10 seconds. Table 3 lists the key parameters of the current imaging system of NVST. Some cases observed recently are summarized as follows (see also Figure \ref{Fig7}): the photosphere bright points, the fine structures of solar activities, the micro solar activities such as micro-filaments and their evolution and so on. These cases are the forefront topics of modern solar physics and need high angular resolution near the diffraction limit of NVST.

%      A figure as large as the width of the column
%-------------------------------------------------------------
%

%               one-column-spanning table
%________________________________________ Table 2: Use_of_the routines
\begin{table}
\begin{center}
\caption[]{ Key parameters of the Multi-Channel High Resolution Imaging System.}
\label{Tab:tab3}

%%Please Capitalize the First Letter of Each Notional Word in table's caption

 \begin{tabular}{llll}
  \hline\noalign{\smallskip}
  Channels & H-alpaha & TiO-band & G-band \\
  \hline\noalign{\smallskip}
     Filters                          & Lyot filter            & Interference filter  & Interference filter \\
     Wavelength({\AA})                & 6,562.8, $\pm$5 Tunable & 7,058	              & 4,300 \\
     Band pass({\AA})                 & 0.25                   & 10                   & 8 \\
     Focal length(m)                  & 22.5	               &35(26)	              & 35 \\
     Frames\/sec(fps)                 & 10                     & 10                   &10  \\
     Field of View $\ast$(arc-second) & 180$\times$180         & 180$\times$180       & 180$\times$180  \\
  \hline\noalign{\smallskip}
     \multicolumn{4}{l}{$\ast$ The FOV size in this table is the maximum area without vignetting. The real value of this }\\
     \multicolumn{4}{l}{parameter is normally limited by the chip size of detector.}\\
  \noalign{\smallskip}\hline
\end{tabular}
\end{center}
\end{table}

\subsection{Spectrometers}

The spectral measurement of the solar atmosphere is another important task of NVST. It is the primary method to detect the magnetic and velocity fields in the solar atmosphere. Besides the classical scientific goals, the small scale structures in the photosphere and the chromosphere could be detected by the spectrometers combined with the AO system. These small structures include bright points, dynamic features of plasma inside a flux tube, fine structures of the quiescent filament, and so on.

\begin{figure}
   \centering
   \includegraphics[width=\textwidth, angle=0]{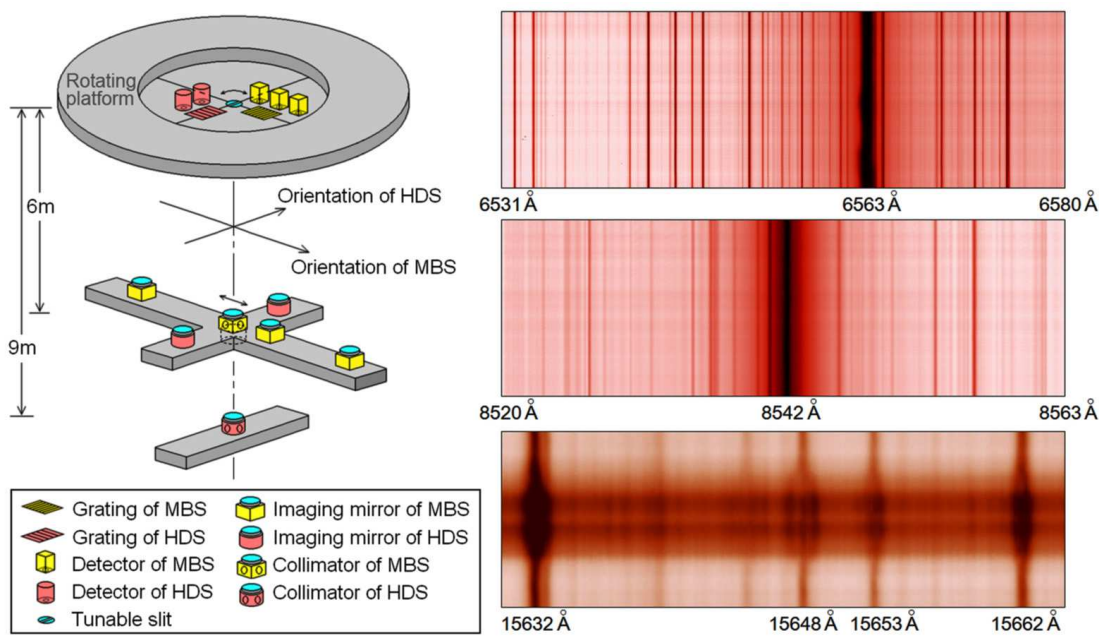}
   \caption{3-D sketch of the spectrometers (right, not in scale), and some observational results (left).The spectra around 6,563 {\AA} and 8,542 {\AA} lines of a quite region were observed by MBS. The spectra around 15,600 {\AA} lines of a small sunspot were observed by HDS.}
   \label{Fig8}
   \end{figure}

The spectrometers of NVST include a multi-bands spectrometer (MBS) and a high dispersion spectrometer (HDS). They are placed on the vertical hanging bracket right below the rotating platform, which rotates along with the platform as a huge image de-rotator (Figure \ref{Fig3}). A tunable slit of the spectrometers is installed at the center of the rotating platform. The width and the orientation of the slit can be changed for different spectrometers and various science cases. Before the slit, as described in previous section (Figure \ref{Fig5}), about 10\% of the photons are transferred into imaging system to display the image and the position of slit. The orientations of the two spectrometers are perpendicular to one another (see left panel of Figure \ref{Fig8}), and they share the same slit described above. Switching use of the spectrometer from MBS to HDS is realized by rotating the slit and moving the collimating mirror of MBS. This operation could not run during observations, and should be completed beforehand. Therefore, the two spectrometers cannot work simultaneously.

The gratings equipped for the two spectrometers are different. We are using a blazed grating for MBS and an echelle grating for HDS. The two gratings are both installed on the reverse side of the rotating platform, and face the collimating mirrors (see Figure \ref{Fig5}). Imaging mirrors and collimating mirrors are placed on the vertical hanging brackets. Detectors are all placed face down on the central part of the rotating platform to record the spectrograms focused by imaging mirrors. The huge space of spectrometers keeps the scattering light at a low level. Important parameters of MBS and HDS are shown in Table 4, and some observational results obtained by the two spectrometers are displayed in the right panel of Figure \ref{Fig8}. These results are preliminary processed and calibrated in order to correct the obvious aberrations and remove the visible interference fringes (\cite{wangr13}).

We note here that the pixel size of each instrument is not listed in Tables 3 and 4, because even the same instrument needs different detectors for different cases. Usually, various detectors, such as the CCD, CMOS and EMCCD have different formats and pixel sizes. So, values of these parameters will be included in the head of the files for observational data, and should be changed from case to case.

%               one-column-spanning table
%________________________________________ Table 2: Use_of_the routines
\begin{table}
\begin{center}
\caption[]{ Key parameters of the spectrometers.}
\label{Tab:tab4}

%%Please Capitalize the First Letter of Each Notional Word in table's caption

 \begin{tabular}{lll}
  \hline\noalign{\smallskip}
  Spectrometer  & MBS & HDS \\
  \hline\noalign{\smallskip}
     Grating(mm$^{-1}$)                      & 1,200 (blazed grating) & 316 (echelle grating) \\
     Blazed angle(degree)                    & 36.8                  & 63 \\
     Focal length of imaging mirrors(m)      & 6	                 & 6 \\
     Focal length of collimating mirror(m)   & 6                     & 9 \\
     Effective size of grating (mm)          & 156$\times$130        & 334$\times$196 \\
     Primary lines({\AA})                    & 5,324, 8,542, 6,563	     & 10,830, 15,648  \\
     Dispersion(mm/{\AA})                    & 0.75 (at 6563 {\AA})  & 0.77 (at 15,648 {\AA}) \\
     Resolution($\lambda$/$\Delta$$\lambda$) & 130,000	             & 300,000$\sim$400,000  \\
     Spectrum range({\AA})                   & $>$ 50	             & $>$ 20  \\
  \noalign{\smallskip}\hline
\end{tabular}
\end{center}
\end{table}

%      A figure as large as the width of the column
%-------------------------------------------------------------

\subsection{Upgrading of instruments}

The instruments described above are currently installed on NVST, but the present situation does not remain unchanged. Upgrading these instruments is always under consideration, and the consequent deployment would follow when it becomes necessary and technically doable. Furthermore, enough space for the new instruments on the 6 m rotating platform has also been left, and any newly developed instruments could be installed when the relevant topics or questions need to be studied. In addition to the platform, the large hanging bracket could accommodate more imaging mirrors as well which allows the spectrometers to observe more lines if necessary.

The ongoing upgrading of instruments involves the AO system and the multi-channel imaging system. First, a high order AO system with 127 actuators is under construction, and will replace the present one as it is done. Then, two new channels with tunable Lyot filters will be added at 10,830 and 3,933 {\AA} bands. After this upgrading, NVST will be able to do more comprehensive chromosphere observations with higher spatial resolution. Besides the running upgrading, more proposals have been put forward. An integral field spectrometer with about 400 fibers is in the progress of development, and its spectral resolution will be up to 100,000. A high spatial resolution magnetograph is also designed in order to match the resolving power of the current multi-channel imaging system. NVST may also be the primary laboratory for the instruments that will be installed on the next generation of solar telescopes, such as the Chinese Giant Solar Telescope (\cite{liu12}). Some typical new technologies and new instruments will be tested and examined on NVST, including the multi-conjugate adaptive optics (MCAO; \cite{becker88}), the two-dimension real-time spectrograph (\cite{ai93}), and the large field fiber array (\cite{dun12}).

\section{High resolution observation by NVST}
\label{sect:High}

The random and rapid image degradation induced by the earth's turbulent atmosphere is the major problem of ground-based telescopes. The image degradations of NVST could not be perfectly corrected by its current low order AO. So, in order to get the high spatial resolution images, no matter with or without AO, the raw data from NVST should be reconstructed by high resolution imaging algorithm. The reduced data from NVST are classified into two levels. The level1 data are processed by frame selection (lucky imaging) (\cite{rob03}). The level1+ data are reconstructed by speckle masking (\cite{weig77}; \cite{lohm83}) or iterative shift \& add (ISA; \cite{liu98}).

\begin{figure}
   \centering
   \includegraphics[width=\textwidth, angle=0]{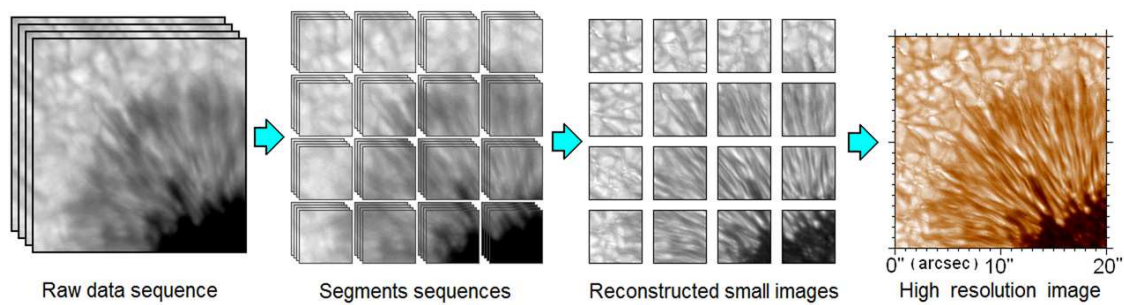}
   \caption{Flow chart of high resolution image reconstruction.}
   \label{Fig9}
   \end{figure}

A high resolution image is normally reconstructed from at least 100 short exposure images. Each raw image is divided into many 5 arc-seconds square segments to match the isoplanatic angle of the earth's atmosphere. Therefore, 100 raw images are divided into a number of subsequences. A high resolution subimage is reconstructed from a subsequence which contains 100 small segments. Finally, combine all the reconstructed subimages to form a whole high resolution image (Figure \ref{Fig9}). The seeing parameter that is needed for reconstruction is calculated by spectrum ratio (\cite{von84}) or measured by SDIMM. The time cadence of a reconstructed image is limited by the observation time of a raw data sequence. The scanning and readout speed of all the detectors on NVST are fast enough to insure the highest time cadence smaller than 10 seconds.

%      A figure as large as the width of the column
%-------------------------------------------------------------

High resolution observation of active region is the primary science task of NVST in 24th solar cycle. After reconstruction, the spatial resolution of photosphere (G-band, 4,300 {\AA}) image could reach near 0.1 arc-second in good seeing condition, and the spatial resolution of chromosphere (H$\alpha$, 6,563 {\AA}) image is better than 0.3 arc-second. With such a spatial resolution, NVST can resolve many fundamental structures of the photosphere and chromosphere within 3 arc-minutes field of view (FOV; \cite{xu14}).

\begin{figure}[b]
   \centering
   \includegraphics[width=\textwidth, angle=0]{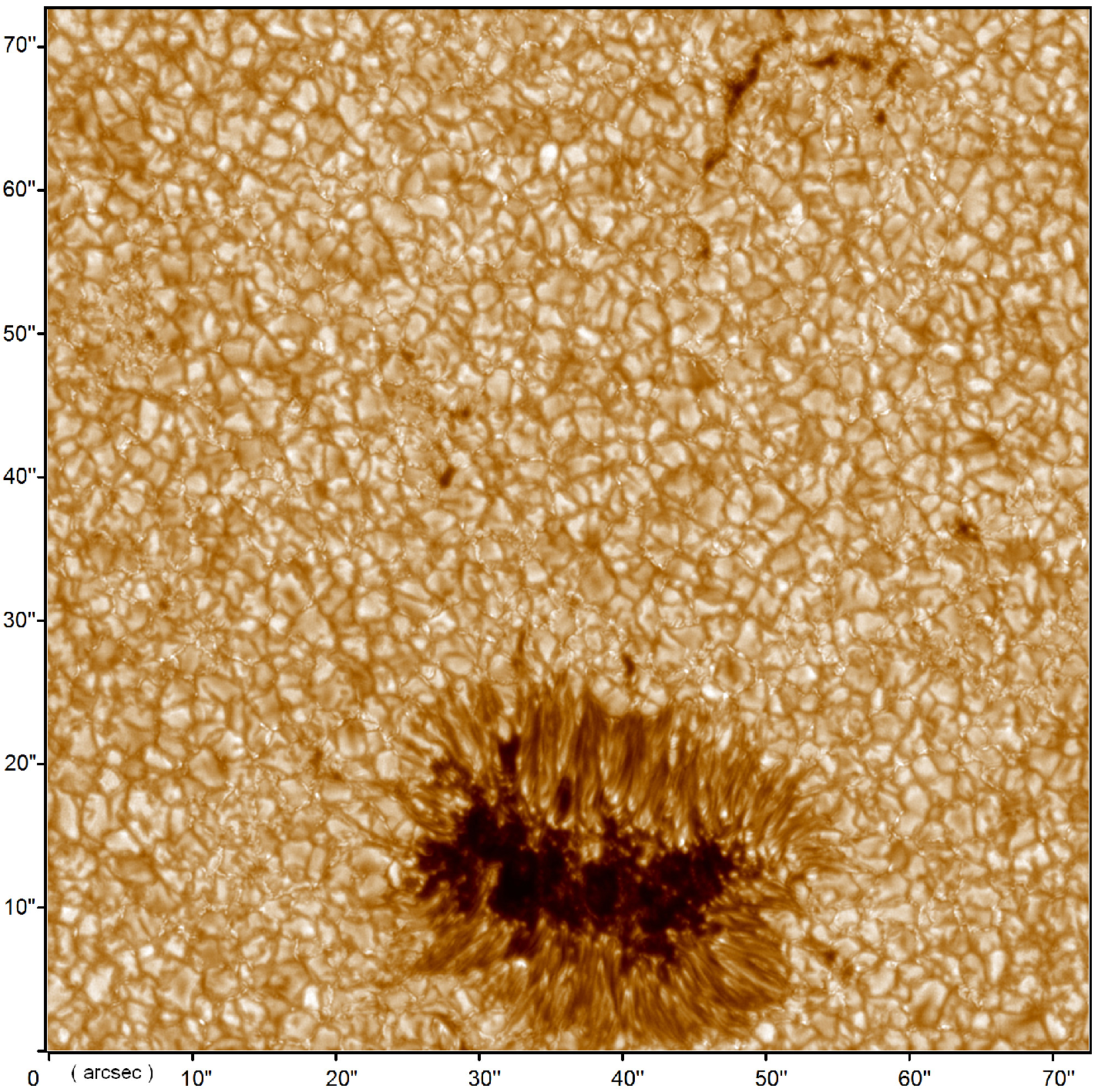}
   \caption{ High resolution photosphere image of AR11154. Level1+ data at 7,058 {\AA}.}
   \label{Fig10}
   \end{figure}

\begin{figure}
   \centering
   \includegraphics[width=\textwidth, angle=0]{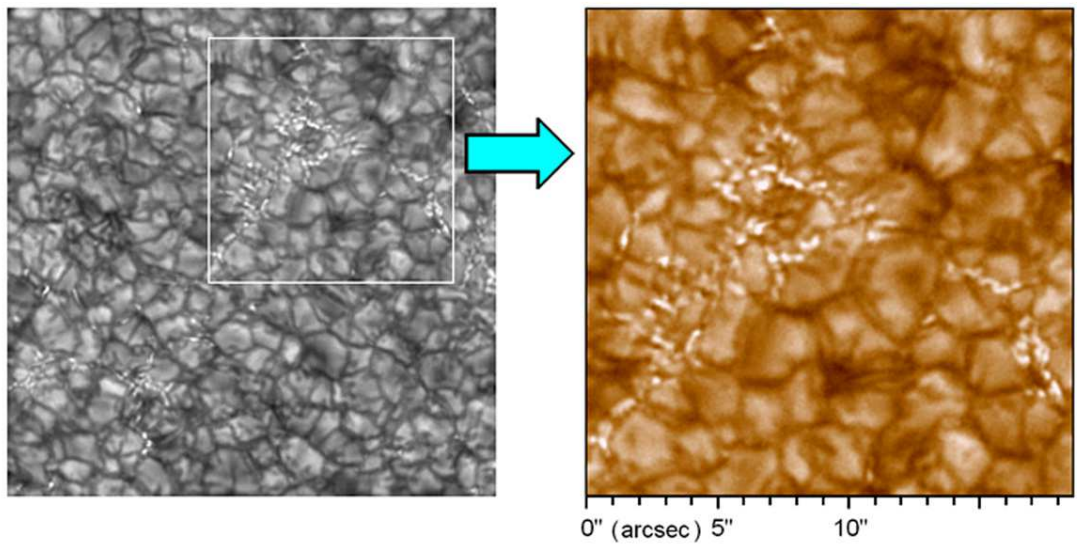}
   \caption{G-band bright points of quite sun. Level1+ data at 4,300 {\AA}.}
   \label{Fig11}
   \end{figure}

\begin{figure}
   \centering
   \includegraphics[width=\textwidth, angle=0]{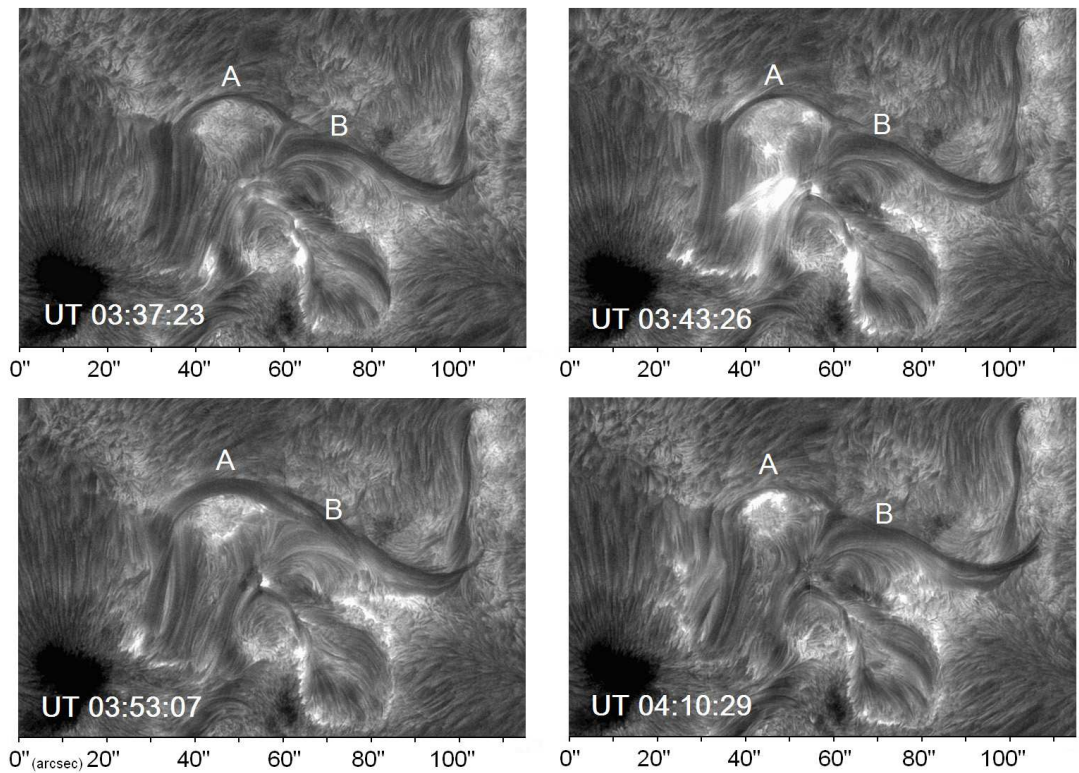}
   \caption{Quick evolutions of the active filaments during a flare eruption on October 25, 2012 (AR11598). Level1+ data at H$\alpha$ line center.}
   \label{Fig12}
   \end{figure}

\begin{figure}
   \centering
   \includegraphics[width=\textwidth, angle=0]{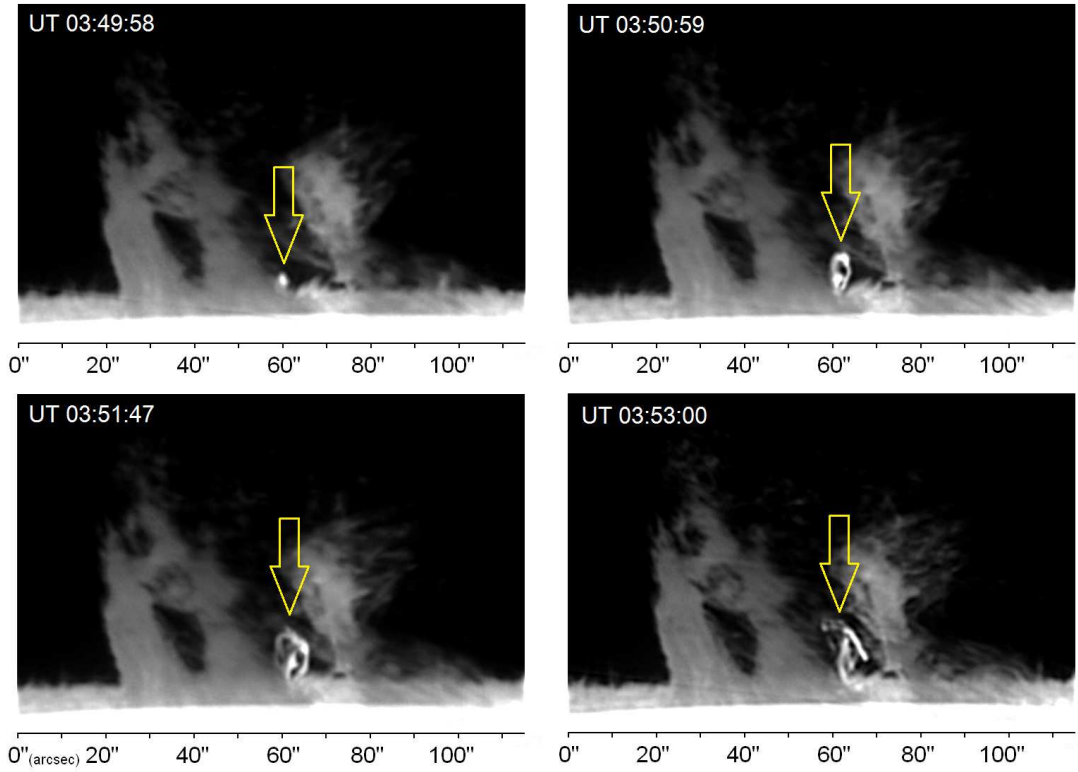}
   \caption{A quiescent prominence and a rapid rising flow on February 15, 2013. Level1 data at H$\alpha$ line center.}
   \label{Fig13}
   \end{figure}

\begin{figure}
   \centering
   \includegraphics[width=\textwidth, angle=0]{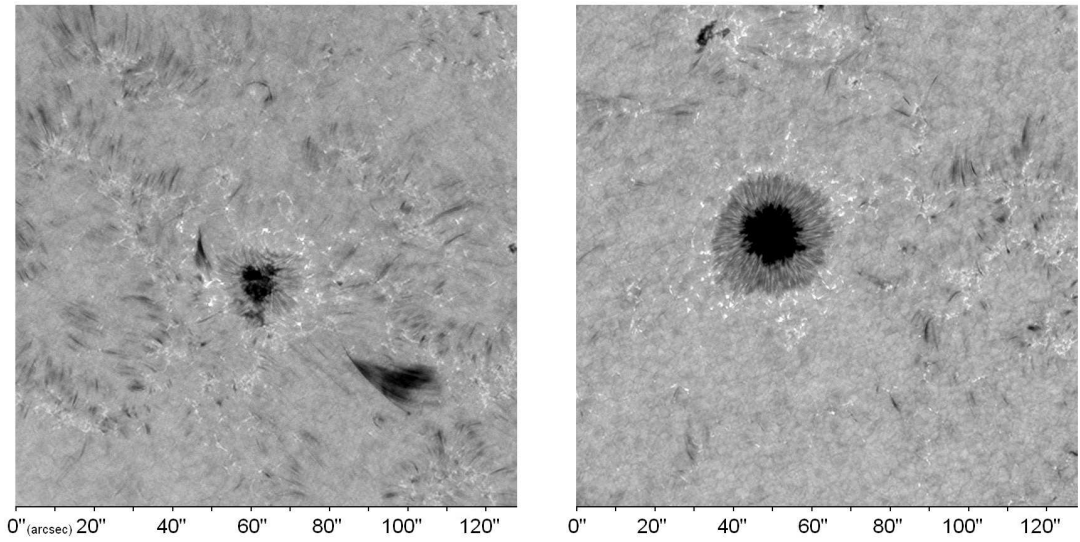}
   \caption{The fine structures around the small sunspots. Level1+ data at H$\alpha$ blue wing (-0.8 {\AA}).}
   \label{Fig14}
   \end{figure}

Figures \ref{Fig10} to \ref{Fig14} are some high resolution reconstructed images of NVST. These results represent the different objects and various science cases. All the results have been reconstructed by the above high resolution imaging algorithms. Figure \ref{Fig10} is a high resolution photosphere image of a typical small active region at TiO band. In this image, the photosphere details such as the umbra bright points, the penumbra fibers and the bright points between granules are all resolved clearly. This kind of data is very useful for researches on the fine structures and evolutions of active regions. Figure \ref{Fig11} is another high resolution photosphere image of quite sun at G-band. It shows the high resolving power of NVST, and the spatial resolution of this image is almost up to the diffraction limit of NVST. Such data can be used to study the dynamical properties of the photosphere fine features and also very important to look for photosphere foot points of the chromospheric activities. Figure \ref{Fig12} includes several high resolution H$\alpha$ images of AR11598 and the surrounding filaments. It shows the quick evolutions of active filaments during a flare eruption. The primary target of Figure \ref{Fig13} is a quiescent prominence at the edge of solar disk. The quick formation process of another small prominence is clearly shown too. Figure \ref{Fig13} is also shows that the system scattering of NVST are low enough to observe the low contrast objects. Figure \ref{Fig14} includes two off-band H$\alpha$ images. Spicules and Ellerman bombs around active regions are the targets of this observation. It should be noted, as the seeing of FSO is normally stable during several hours, the high resolution data sequences of NVST are usually long enough to cover most of the short term solar activities, such as a classical flare and the process of a filament eruption.

\section{Summary}
\label{sect:suma}
The observation data of NVST including images and movies have been opened at present (http://fso.ynao.ac.cn). As one of the current big solar telescopes in the world, NVST shows the expected power in the high resolution observations. Considering its location is just between Europe and America, NVST could combine with the other big solar telescopes (\cite{scha03}; \cite{schm12}; \cite{good12}) to form a global high resolution observation net. NVST is expected to contribute some original discovers in the near future, so it is necessary to equip more high-precision instruments, especially the instruments of magnetic-field measuring.

\begin{acknowledgements}
NVST was firstly proposed by professor Guoxiang Ai and professor Cheng Fang. Professor Jingxiu Wang and professor Jinxin Hao gave us the most important support in the early stage of this project. The Authors would like to give the great appreciate to professor Jacques M. Beckers for his valuable suggestions about the telescope and his contributions in site testing. Professor Haisheng Ji and Professor Jun Lin gave us many good suggestions to improve this paper. Thanks so much to all NVST team members for their outstanding works.

NVST is supported by the whole Chinese solar community. Besides the institutes of all authors, the important contributors should include Nanjing University, NAIRC of CAS and LZOS of Russia. The manufacture of the telescope and instruments were funded by the National Key Basic Research Program of China (2011CB811400, 2006CB806300, G200078400), the CAS (KJCX2-EW-T07) and the NSFC (11178004, 11103077, 11273058, etc.).
\end{acknowledgements}

%      A figure as large as the width of the column
%-------------------------------------------------------------
%

%      A figure as large as the width of the column
%-------------------------------------------------------------
%

%      A figure as large as the width of the column
%-------------------------------------------------------------
%

%      A figure as large as the width of the column
%-------------------------------------------------------------
%

%      A figure as large as the width of the column
%-------------------------------------------------------------

\label{lastpage}

\end{document}